\documentstyle[12pt]{article}
\normalbaselines
\catcode `\@=11
\@addtoreset{equation}{section}

\def\theequation{\arabic{section}.\arabic{equation}}

\def\section{\@startsection {section}{1}{\z@}{-3.5ex plus -1ex minus
     -.2ex}{2.3ex plus .2ex}{\normalsize\bf}}
\def\subsection{\@startsection{subsection}{2}{\z@}{-3.25ex plus -1ex minus
 -.2ex}{1.5ex plus .2ex}{\normalsize\bf}}

\def\thebibliography#1{\section*{References\markboth
  {REFERENCES}{REFERENCES}}\list
  {[\arabic{enumi}]}{\settowidth\labelwidth{[#1]}\leftmargin\labelwidth
  \advance\leftmargin\labelsep
  \usecounter{enumi}}
  \def\newblock{\hskip .11em plus .33em minus -.07em}
  \sloppy
  \sfcode`\.=1000\relax}
 
\catcode `\@=12

\newcommand{\pl}{\partial_}

\newcommand{\bqq}{\begin{equation}\label}
\newcommand{\eeq}{\end{equation}}

\newcommand{\ad}{\,{\rm ad}\,}

\newcommand{\rr}{{\mbox{\bf R}}}
\newcommand{\cc}{{\mbox{\bf C}}}
\newcommand{\rrr}{{\mbox{\bf \small R}}}
\newcommand{\U}{{\mbox{\tiny U}}}
\newcommand{\qq}{{\cal Q}}

\begin{document}

\vspace*{2.5cm}
\noindent
{\bf ON QUANTIZATION OF POLYNOMIAL
     MOMENTUM OBSERVABLES\footnote{
The work was partially supported
by grants of Kazan Mathematical
Society and Russian Foundation for Basic
Researches (Project No~96-01-01031).}
}\vspace{1.3cm}\\
\noindent
\hspace*{1in}
\begin{minipage}{13cm}
Dmitry A. Kalinin \vspace{0.3cm}\\
Department of General Relativity and Gravitation,\\
Kazan State University, 18 Kremlyovskaya Str.,\\
Kazan 420008 Russia\\
e-mail: dmitry.kalinin@ksu.ru
\end{minipage}

\vspace*{2cm}

\begin{abstract}
\noindent
The paper is devoted to quantization of polynomial momentum observables
in the cotangent bundle of a smooth manifold. 
A quantization procedure is proposed allowing
to quantize a wide class of functions which are polynomials
of any order in momenta. In the last part of the paper
the quantum mechanics of scalar particle in curved space-time
is studied with the use of proposed approach. 
\end{abstract}

\vspace*{1cm}

\section*{INTRODUCTION}

In the paper, the problem of quantization of polynomial 
classical observables is studied. 
It is well-known that it is impossible to fully quantize
the algebra of polynomials on a Euclidean phase space
\cite{vanhove,hurt,GotGrTuy}. Consequently, research has concentrated in
two main directions. The first is weakening the notion of ``full"
quantization. In this direction various quantization schemes
have been developed to the present time \cite{GotGrTuy,kir2,BFFLS,KM}.

The second direction is quantizing of restricted classes of observables.
Such classes are usually connected with symmetries of the phase
space \cite{sniat1,amka1}. If we are interested in phase spaces $N$
which have a distinguished configuration space, i.e. $N=T^* M$, then
an important class of classical observables is one of functions
polynomial in momenta. The results obtained previously in quantizing
of such functions deals mainly with observables quadratic in
momenta {sniat1,gotay}.

In present paper, a quantization procedure is proposed allowing
to quantize a wide class of functions which are polynomials
of any order in momenta. We use the geometric quantization procedure
of Kirillov, Kostant and Souriau (see \cite{kir2,sniat1,Woodhouse} for
review). As an application of proposed procedure, the quantization
of the Hamiltonian of a scalar pointlike
particle in curved space-time is considered.

The plan of the paper is as follows. In Sec.~1 the main aspects of
geometric quantization scheme are presented. Next section is devoted to the
quantization of polynomial momentum observables
in a phase space which is the cotangent bundle of a smooth
manifold. In the last part of the paper
the quantum mechanics of scalar particle in curved space-time
is studied with the use of the proposed approach. 

\section{\hspace{-4mm}.\hspace{2mm}GEOMETRIC QUANTIZATION}

In this section we shortly describe the geometric quantization
method which will be used in following sections (see 
\cite{kir2,sniat1,Woodhouse} for review).

Let $(N,\omega)$ is a symplectic manifold and $C^\infty  (N)$ be the set of
$C^\infty$-functions on $N$. For any $f\in C^\infty (N)$ 
its {\it Hamiltonian vector field} $\ad (f)$ can be defined
by the equation $\iota (\ad (f)) \omega = -df,$
where $\iota$ is the internal product. The {\it Poisson bracket}
    $$
\{f_1,f_2 \} = \ad (f_1) (f_2), \quad f_1,f_2 \in C^\infty (N)
    $$
turns $C^\infty (N)$ into a Lie algebra -- the 
{\it Poisson  algebra  of} $N$. If the Hamiltonian vector
field $\ad (f)$ is complete then $f$
generates via the Poisson brackets a one-parameter group $\phi^t_f$
of canonical transformation of $N$ \cite{arnold,kalinin}.

A {\it quantization} is the linear map ${\cal Q}$ of
a subalgebra ${\cal F}\subset 
C^\infty (N)$  into the set of self-adjoint operators in
a Hilbert space ${\cal H}$ possessing the following
{\it quantization axioms}

(Q1) ${\cal Q}\; 1 = {\rm id}_{\cal H}$;

(Q2) $[{\cal Q}\; f_1, {\cal Q}\; f_2 ] = i\hbar {\cal Q}\;
(\{ f_1 , f_2 \})$ where $\hbar=h / 2\pi$ is the Planck constant;

(Q3) for a complete set of functions
$f_1,\ldots,f_r$ the operators ${\cal Q} (f_1) ,\ldots,
{\cal Q} (f_r)$ also

form a complete set \cite{kir2}.

{\it Geometric quantization} is an explicit
realization of the quantization map (see \cite{BFFLS,KM}
for review of another quantization approaches) providing
a constructive way of quantizing subalgebras of the Poisson
algebra.

The geometric quantization procedure includes the following
main components \cite{kir2,hurt,sniat1,Woodhouse}.

a) {\it Prequantization line bundle} ${\cal L}$ that is a Hermitian
line bundle over $N$ with connection $D$,
$D$-in\-va\-ri\-ant\footnote{Recall that $D$-invariance means that for
each pair of sections $\lambda$ and $\mu$ of ${\cal L}$ and each real
vector field $X$ on $M$ holds $X<\lambda ,\mu >=<D_X \lambda
,\mu>+<\lambda,D_X \mu>$.} Hermitian structure $<,>$. The connection form
$\alpha$ of $D$ should obey the prequantization condition
$d\alpha = -h^{-1}\omega$.

b) {\it Polarization} $F$ that is an integrable Lagrange distribution
in $TN\otimes_{\rrr}{\cc}$. The polarization plays an important role in
the geometric quantization approach, because it determines the
representation space.

c) {\it Metaplectic structure} which consists of the {\it bundle of
metalinear frames } and the {\it bundle ${\cal L}\otimes
\sqrt{\land^n F}$ of ${\cal L}$-va\-lu\-ed half-forms normal to the
polarization $F$}.

If these structures are defined on a symplectic manifold
($N,\omega$), then the Hilbert space ${\cal H}$ is defined to be the
space of sections $\mu$ of the bundle ${\cal L}\otimes 
\sqrt{\land^n F}$ which are covariantly constant along the 
polarization $F$. A function $f\in C^\infty (N)$ is
called {\it polarization preserving} if its Lie derivative
$L_X \ad (f) \in F$ for any vector $X\in F$. For such functions the
quantization is defined by the {\it Souriau-Kostant prequantization formula}
    \bqq{SKQ}
{\cal Q} (f) \sigma = (f-i\hbar D_{\ad (f)}) \sigma,\qquad
D_X \sigma =0, \quad \sigma\in {\cal H},\; X\in F.
    \eeq
In order to quantize functions not preserving polarization one should
construct a mapping connecting the Hilbert spaces ${\cal H}_F$
corresponding to different polarizations. A way of doing it is provided
by the Blat\-t\-ner-Kos\-tant-Ster\-n\-berg (BKS) kernel
\cite{sniat1,Woodhouse}.

Further we shall consider a particular case when the system under
consideration has a physically distinguished {\it configuration space}
$M$, i.e. $N=T^* M$. Define the {\it canonical} 1-{\it form} $\theta$ by
$\theta (u) = \xi (\pi_* (u))$ for each $\xi \in T^* M$,
$u\in T_\xi \; T^* M$ where $\pi: T^* M\to M$ is the cotangent bundle
projection and $\pi_*$ is its differential. The 2-form $\omega=d\theta$
is closed and non-degenerate and turns $T^* M$ into a symplectic manifold. 

For each smooth vector we define a function $P_X$ on $T^* M$ by
$P_X (\xi) = \xi (X (\pi (\xi)))$, $\xi \in T^* M.$
We shall refer to $P_X$ as the {\it momentum associated to a vector 
field} $X$.

If $x^\alpha$, $\alpha=1,\ldots,n$ are local coordinates in a domain
$U\subset M$, then we can introduce the {\it canonical coordinate system}
$(\pi^{-1} (U), x^\alpha, p_\alpha)$ where $p_\alpha = P (\partial /
\partial x^\alpha)$. In these coordinates on $T^* M$ the symplectic form
$\omega$ takes the canonical form $\omega=\sum_\alpha dp^\alpha \land
dx_\alpha$. Let $\pi: TM \to M$ be the tangent bundle projection, then 
the {\it vertical polarization} $F=$Ker$\; \pi$ can be defined spanned
by the vector fields $\partial /\partial x^\alpha$ .

It is easy to demonstrate that the vertical polarization leads to the
Schr\"o\-din\-ger representation. For construction of the Hilbert space
${\cal H}_F$ corresponding to this representation we should define a global
section of the bundle of metalinear frames ${\cal L}$ $\otimes$
$\sqrt{\land^n F}$ over $T^* M$. We demonstrate further
(see \cite{sniat1} for details) how it can be
done with the use of a non-degenerate symmetric tensor field $\gamma$ of
second rank on the configuration space $M$.
This tensor field can be considered
as a (pse\-u\-do) Riemannian metric on $M$.
              
First, we note that there exists an isomorphism between the sections
$\mu$ of the bundle ${\bf B}\equiv \sqrt{\land^n F}$ and the set of
complex-valued functions $\mu^{\#}$ on ${\bf B}^*\equiv {\bf B}\backslash
\{ 0 \}$ possessing the property $\mu^{\#} (cz)=c^{-1} \mu^{\#} (z),
\quad c\in \cc.$ This isomorphism is given by $\mu (\pi z)= \mu^{\#} (z)z$
where $\pi:{\bf B}\to T^* M$ is the projection.

For each chart $(U,x^\alpha)$ on $M$, the metric $\gamma$
defines a matrix-valued function $\gamma_\U=(\gamma_{\U \alpha\beta})$ on
$U$. Let $\zeta_\U=\{ \zeta_1, \ldots , \zeta_n \}$ be an orthonormal
frame for $TU$
   $$
\zeta_\alpha=\sum_\beta C^\beta_\alpha
\frac{\partial}{\partial x^\alpha}, \quad \det C >0, \quad
(C \gamma_\U C^T)_{\alpha\beta}=\delta_{\alpha\beta}.
  $$
From here $\det \;C=|\det \gamma_\U |^{1/2}$. Let $\eta_\U
=\{ \eta_1 , \ldots , \eta_n \}$ be the frame field for $F$ defined by
$\eta_\U=\ad (x) (C^T\circ \pi)^{-1}$ and let $\tilde\eta_\U$ be a
metalinear frame field \cite{sniat1} projecting onto $\eta_\U$.
Then we can define a local section $\mu_{\gamma_U}$ of ${\bf B}$ by 
$\mu_{\gamma_U}^{\#}\circ \eta=1$ satisfying
   $$
\mu_{\gamma_U}= \pm |(\det\gamma_\U)\circ \pi|^{1/4}\mu_\U
   $$
where $\mu_\U$ is the metalinear frame field for $\zeta_\U$.
Using the transformation properties of $\gamma$ it can be shown that 
$\mu_{\gamma_U}$ defines a covariantly constant global section
$\mu_\gamma$
of ${\bf B}$ such that, for each open domain $U\subset M$,
   \bqq{mu0}
\mu_\gamma |_{\pi^{-1} (U)} = \pm |(\det \gamma_\U )
\circ \pi |^{1/4} \mu_\U.
   \eeq

If $F$ is the vertical polarization, then the symplectic form is exact
$\omega=d\theta$ and each section $\sigma$ covariantly constant
along $F$ can be represented in the form
   \bqq{vert1}
\sigma = \Psi \: \lambda_0 \otimes \mu_\gamma
   \eeq
where $\lambda_0$ is a non-vanishing section of ${\cal L}$ and $\Psi$ is 
a complex-valued function on $M$.

The most general quantization formula for a function $f\in C^\infty (N)$
by the geometric quantization procedure can be written in the
form \cite{kir2,sniat1}
   \bqq{q}
{\cal Q} (f) \sigma = i\hbar \frac{d}{dt}
(\tilde\phi^t_f\sigma)|_{t=0},\qquad \sigma\in {\cal H}
   \eeq
where $\tilde\phi_f^t$ is the one-pa\-ra\-me\-ter group of transformation
in the Hilbert space ${\cal H}$ generated by the function $f$.
If the support of $\Psi$ is contained in some coordinate
neighborhood $U$, then we can write
   \bqq{2a}
\Psi\; \lambda_0\otimes\mu_\gamma =
\psi (x)\; \lambda_0\otimes \mu_\U
   \eeq
where $\psi$ is a function on $\rr^n$ with support contained in
the image of the chart $x: U\to \rr^n$. Comparing (\ref{2a}) and (\ref{mu0})
we find, for each $y\in U$,
   \bqq{psi}
\psi (x (y)) =\pm \Psi (y) |\det \gamma_\U (y) |^{1/4}.
   \eeq
For sufficiently small $t$
   \bqq{phi-t}
\tilde\phi^t_f (\Psi \; \lambda_0 \otimes \mu_\gamma) =
\psi_t (x)\; \lambda_0 \otimes \mu_\U
   \eeq
where $\psi_t$ is given by the equation \cite{sniat1}
   $$
\psi_t (x)=(i\hbar)^{-n/2}\int [\det \omega (\ad
(x^\mu), \phi^t_f \ad (x^\nu))]^{1/2}
   $$
   \bqq{p-t}
\exp [i/\hbar \int^t_0 (\theta (\ad (f))-f)\circ \phi^{-s}_f ds]
\; \psi (x \circ \phi^{-t}_f) \: d^n p.
 \eeq

\section{\hspace{-4mm}.\hspace{2mm}
QUANTIZATION OF POLYNOMIAL MOMENTUM OBSERVABLES}

In this section we consider the quantization of polynomial 
momentum observables in the phase space
$N=T^* M$ using the geometric
quantization procedure in the Schr\"o\-din\-ger representation.
Consider a classical system with an oriented configuration space
$M$ and let $F$ be the vertical polarization on the phase space $T^* M$.

Let us denote $C_r (N)$ the subspace $C^\infty (N)$ consisting of
the functions which are at most $r$-th degree polynomials along the
fibers of the cotangent bundle projection.
A function $f\in C_r (M)$ can be represented as a sum
of homogeneous terms of the form
    \bqq{monom}
f_{(k)} = f_{(k)}^{\alpha_1\ldots \alpha_k} (x)
p_{\alpha_1}\ldots p_{\alpha_k}, \quad k=1,\ldots,r.
    \eeq
Along with the functions $f_{(k)}$ we shall consider further 
the associated tensor fields $\varphi_{(k)}=
f_{(k)}^{\alpha_1 \ldots \alpha_k}
\partial_{\alpha_1} \ldots \partial_{\alpha_k}$.

In order to construct the quantization map for a function
$f\in C_r (N)$ we should first consider the quantization for
the homogeneous terms (\ref{monom}) of any order in momenta.

At first, let us define a global non-va\-ni\-sh\-ing section $\sigma_0$
of the bundle ${\cal L}\otimes\sqrt{\land^n
F}$ of ${\cal L}$-va\-lu\-ed half-forms normal to the (vertical)
polarization. As it was shown in Sec.~2, such section can
be constructed using a (pseudo)Riemannian metric
$\gamma$ on the configuration space $M$.

In fact we can choose {\it any} such metric on $M$ in order to construct the
section $\sigma_0$. However, in applications we are usually
interesting in a {\it natural choice} of $\gamma$.

The problem simplifies if for given function $f$
the matrix $(f_{(2)}^{\alpha\beta})$ in (\ref{monom})
is non-degenerate. In this case we can put
$\gamma = f_{(2)}{}_{\alpha\beta}\: dq^\alpha dq^\beta$,
$f_{(2)}{}_{\alpha\beta}f_{(2)}^{\beta\tau}=
\delta^\tau_\alpha$. However, if
$(f_{(2)}^{\alpha\beta})$ is degenerate, one should choose
$\sigma_0$ using additional considerations. For example, a
(pseudo)Riemannian structure on the configuration
space $M$ could be used, motivated by some physical reasons.

When the global section of ${\cal L}\otimes\sqrt{\land^n F}$
is chosen, we have to develop an approach for quantization of
the homogeneous terms (\ref{monom}) for any order $k$.
In fact, we propose here a way of quantizing only a wide
class of such terms.

For $k=0,1$ the functions are vertical
polarization preserving and it is easy to write down the expressions
for corresponding operators \cite{sniat1} using
the Kostant-Souriau prequantization formula (\ref{SKQ})
     $$
{\cal Q}( f_{(0)}+f_{(1)}^\alpha
p_\alpha)\: \psi\: \lambda_0\otimes\mu_\gamma=
     $$
     $$
(-i\hbar \ad
(f_{(0)}+f_{(1)}^\alpha
p_\alpha )\: \psi+(f_{(0)}-\frac{i\hbar}{2} \frac{\partial
f_{(1)}^\alpha}{\partial q^\alpha} )\: \psi) \:
\lambda_0 \otimes \mu_\gamma.
     $$

Consider now the case $k\ge 2$. Let us suppose that it is possible
to introduce in a neighborhood of any point $y \in M$ such coordinates
$(x^\alpha)$ that the first partial derivatives of the functions
$f_{(k)}^{\alpha_1 \alpha_2 \ldots \alpha_k}$ at $y$ are equal to zero.
We shall call such coordinates {\it normal with respect to the tensor field}
$\varphi_{(k)}=f_{(k)}^{\alpha_1 \ldots \alpha_k} \partial_{\alpha_1}
\ldots \partial_{\alpha_k}$. The problem of finding the normal coordinates
is equivalent to determining of a (local) linear connection in which the
tensor field $\varphi_{(k)}$ is covariantly constant.

It is easy to see that in general case it is impossible to find such
connection, because the corresponding system of algebraic equations
on connection coefficients is overdetermined. However, the coordinates
normal with respect to a tensor field can be introduced in many important
cases. For example, if $M$ is a (pseudo)Riemannian manifold with the metric
$g$ then it is possible to find the coordinates normal with respect to
covariantly constant tensors $\varphi_{(k)}$ coincides with the Riemannian
normal coordinates. In the next section we
shall see how it allows us to quantize the Hamiltonian of the relativistic
particle on curved space-time background.

In the following we suppose that the function $f_{(k)}$ is such that
in a neighborhood of any point $y \in M$ it is possible to introduce the
coordinates normal with respect to a given tensor field $\varphi_{(k)}$.

Let us now quantize the function $f=f_{(k)} = f_{(k)}^{\alpha_1\ldots
\alpha_k} p_{\alpha_1}\ldots p_{\alpha_k}$ using (\ref{q}) and
(\ref{p-t}). In this case the integral in (\ref{p-t}) can be simplified
if the coordinates $(x^\alpha)$ are normal at the point $y$ with respect
to the tensor field $\varphi_{(k)}$. In this case the functions $x^\alpha$
depend linearly on $t$ along the orbits of the group $\phi^t_f$
   $$
x^\alpha \circ \phi^t_f = kt\; f_{(k)}^{\alpha_1\ldots \alpha_{k-1}
\alpha} p_{\alpha_1}\ldots p_{\alpha_{k-1}}
   $$
for each $\xi \in T^*_y M$. At the same time $p_\alpha \circ \phi^t_f =$const
in these coordinates.

Using the last equation and the formulae
  $$
\frac{d}{dt} (x^\alpha \circ \phi^t_f)=
kf_{(k)}^{\alpha_1\ldots \alpha_{k-1}}
p_{\alpha_1}\ldots p_{\alpha_{k-1}},
  $$
  $$
\frac{d}{dt} (p_\alpha \circ \phi^t_f)=
-\pl{\alpha} f_{(k)}^{\alpha_1\ldots \alpha_k}
p_{\alpha_1}\ldots p_{\alpha_k}
  $$
it is possible to approximate the integrand in (\ref{p-t})
so that the integration will give results accurate to first order
in $t$. After that Eq.~(\ref{p-t}) takes the form
   $$
\psi_t (0)=(i\hbar/k)^{-n/2} \int [\det (f_{(k)}^{\alpha_1\ldots
\alpha_{k-2} \beta \tau}(0))]^{1/2}
   $$
   \bqq{appr-int}
\exp [i/\hbar \; t (k-1) f_{(k)}(0)]
\psi (-kt\; f_{(k)}^{\alpha_1\ldots \alpha_{k-1}
\alpha} (0) \: p_{\alpha_1}\ldots p_{\alpha_{k-1}})) \: d^n p.
   \eeq

It is easy to see that together with (\ref{q}) and (\ref{phi-t})
this formula defines a quantization coinciding with the quantization
of the function
    $$
f_{(k)}^{\alpha_1 \ldots \alpha_k}(0)\:
p_{\alpha_1} \ldots p_{\alpha_k}.
    $$

This function can be quantized using the {\it von Neumann rule}
(see Appendix)
    \bqq{vN}
{\cal Q} (a^{\alpha_1 \ldots \alpha_k}
p_{\alpha_1} \ldots p_{\alpha_k}) =
a^{\alpha_1 \ldots \alpha_k} { \cal Q} (p_{\alpha_1})
\ldots {\cal Q} (p_{\alpha_k}),
    \eeq
    $$
a^{\alpha_1 \ldots \alpha_k}={\rm const}.
    $$
From here using (\ref{vert1}) we get
    \bqq{quant-fk}
{\cal Q} (f_{(k)}) \psi \: \lambda_0\otimes\mu_\gamma=
(i \hbar)^k f_{(k)}^{\alpha_1 \ldots \alpha_k} (0) \:
\frac{\partial^k
(|g|^{1/4}\psi )}{\partial{x^{\alpha_1}} \ldots
\partial{x^{\alpha_k}}}\:(0)\: \lambda_0\otimes\mu_\gamma.
    \eeq

Because the point $y\in M$ in these considerations was chosen
to be appropriate, the formula (\ref{quant-fk}) give the quantization
operator corresponding to any homogeneous term of the form (\ref{monom}).

\section{\hspace{-4mm}.\hspace{2mm}
QUANTUM MECHANICS IN CURVED SPACE-TIME}

As an application of the approach proposed in the previous section
let us consider quantization of Hamiltonian of a particle in
a curved space-time which is a general Riemannian manifold $({\cal M},g)$.

We develop here the quantum theory of a pointlike scalar particle
in curved space-time as
quantization of its classical mechanics. This approach to construction
of quantum theory is almost obvious, however it is badly developed
in relativistic case because of well-known difficulties with quantizing of
non-linear observables. Usually, the field theoretical
approach to construction of quantum mechanics in curved space-time
has been used instead \cite{gorb,tagirov} which deduces the
(finite-dimensional) quantum mechanical description from
the field equations describing corresponding
in\-fi\-ni\-te-di\-men\-si\-o\-nal field systems.
\medskip

Consider first the formulation of classical mechanics of the system.
It can be formulated
by the use of an instant spacelike surface $\Sigma\subset {\cal M}$
\cite{qm-cs1}.
If we choose the local
coordinates $v^\alpha$, $\alpha=1,\ldots,3$
on $\Sigma$ then the Hamiltonian of the
particle takes the form 
    $$
H = \sqrt{\rho^{\alpha\beta}p_\alpha p_\beta + g_{00} \: m^2}
    $$
where $m$  is the particle's mass and 
$\rho$ is the induced metric on $\Sigma$, $\rho_{\alpha\beta} =
g_{\alpha\beta}/g_{00}$.

Let us consider approximation of $H$ for small
values of momenta $\rho^{\alpha\beta}p_\alpha p_\beta << m^2$
   \bqq{appr}
H=H_1+H_2+H_3+\mbox{higher order terms}.
   \eeq
where
   $$
H_1=m\sqrt{g_{00}},\qquad
H_2=\sqrt{g_{00}}\;\frac{g^{\alpha\beta}p_\alpha p_\beta}{2m}, \qquad
H_3=m\sqrt{g_{00}}\;\frac{(g^{\alpha\beta}p_\alpha p_\beta)^2}{4 m^3}.
   $$
For simplicity further we consider only the case $g_{00}=1$ whence 
$\rho_{\alpha\beta}=g_{\alpha\beta}$. Quantizing
the first two terms in this expression we get \cite{sniat1,dewitt}
   \bqq{ftt}
m -\frac{\hbar^2}{2m}g^{\alpha\beta}
\nabla_\alpha \nabla_\beta +\frac{\hbar^2}{12m} R
    \eeq
where $\nabla_\alpha$ is the covariant derivative with respect to
3-metric $\rho$ and $R$ is the scalar curvature.

Let us consider the third term in (\ref{appr}). Introducing the 
Riemannian normal coordinates and applying the method
developed in the previous section
we see that quantization of the function 
$(g^{\alpha\beta}p_\alpha p_\beta)^2$ leads in these coordinates to
the quantization of the function $\sum (p_\alpha)^4$. From here
     \bqq{h3}
{\cal Q}\:(H_3)\:\psi (x)=\frac{\hbar^4}{4m^2} \sum_{\alpha}
\frac{\partial^4 \psi (x)}{\partial v_\alpha^4} (0).
     \eeq
Taking in account the equation (\ref{psi}) we can find the expression
for $({\cal Q}H_3)\psi$ in an appropriate coordinates. In order to do it
we first list all possible independent invariant terms which could appear
in this expression:
     $$
I_1=R^2\psi,\quad I_2=\bigtriangleup R\psi, \quad
I_3= R^{\alpha\beta\mu\nu} R_{\alpha\beta\mu\nu}\psi,
     $$
     $$
I_4= R^{\alpha\beta} R_{\alpha\beta},\quad I_5=
R^\alpha\psi_{,\alpha},\quad I_6 = R^{\alpha\beta}
\psi_{,\alpha\beta},\quad
     $$
     $$
I_7= R\bigtriangleup \psi,\quad I_8=g^{\alpha\beta}
g^{\mu\nu} \psi_{,\alpha\beta\mu\nu}.
     $$
Here the comma denotes the covariant derivative $\nabla_\alpha$,
$R_{\alpha\beta\mu\nu}$, $R_{\alpha\beta}$ and $R$ are Riemann, Ricci
and scalar curvatures of $\rho$ and $\bigtriangleup= g^{\alpha\beta}
\nabla_\alpha \nabla_\beta$. Using (\ref{psi}), in Riemannian normal
coordinates we find from (\ref{h3})
     $$
g^{\alpha\beta} g^{\mu\nu} \pl{\alpha\beta\mu\nu}(|g|^{1/4}\psi)=
g^{\alpha\beta} g^{\mu\nu} (|g|^{-3/4} \pl{\beta\mu}|g|
\pl{\alpha\lambda}\psi +|g|^{-3/4}\pl{\beta\lambda\mu}|g|\pl\alpha\psi
+
     $$
     \bqq{main}
\frac{1}{2}|g|^{-3/4}\pl{\lambda\mu}|g|\pl{\alpha\beta}\psi +
(-\frac{3}{8}|g|^{-7/4}\pl{\alpha\lambda}|g|\pl{\beta\mu}\psi -
\frac{3}{16}|g|^{-7/4}\pl{\alpha\beta}|g|\pl{\lambda\mu}\psi
     \eeq
     $$
+\frac{1}{4}|g|^{-3/4}\pl{\alpha\beta\lambda\mu}|g|)\psi
+|g|^{-1/4}\pl{\alpha\beta\lambda\mu}\psi)
     $$
where $|g|=\det (g_{\alpha\beta})$.

Using the expansion for the components of the metric $g$ in
Riemannian normal coordinates \cite{petrov}
     $$
g_{\mu\nu}=g_{\mu\nu}(0)+ \frac{1}{3}R_{\mu\alpha\nu\beta}(0)v^\alpha
v^\beta - \frac{1}{6}R_{\mu\alpha\nu\beta,\gamma}(0)v^\alpha
v^\beta v^\gamma +
     $$
     $$
(\frac{1}{20}R_{\mu\alpha\nu\beta,\gamma\lambda}(0)+
\frac{2}{45}R_{\alpha\mu\beta\sigma} R^\sigma_{\gamma\nu\lambda}(0))
v^\alpha v^\beta v^\gamma v^\lambda + \mbox{higher order terms},
     $$
Using this formula, after complicated but straightforward
calculations, we get from (\ref{h3}) and (\ref{main})
     $$
{\cal Q}\: (H_3)\: \psi\: \lambda_0\otimes\mu_\rho=
|g|^{1/4}(-\frac{1}{12}I_1+ \frac{1}{5}I_2+ \frac{3}{10}I_3+
\frac{1}{30}I_4- \frac{1}{3}I_5+ \frac{1}{3}I_7+
I_8)\lambda_0\otimes\mu_\rho.
     $$

By the use of the proposed approach it is in principle possible
to calculate quantum operators corresponding to all terms in the
expansion (\ref{appr}) of the particle Hamiltonian. In
order to do it, for example, for the term $(g^{\alpha\beta}p_\alpha
p_\beta)^3$ one should calculate all possible invariants of 6{\it th}
order in normal coordinates and, then rewrite the expression
$g^{\alpha\beta} g^{\lambda\mu} g^{\nu\rho} \partial_{\alpha\beta\lambda
\mu\nu\rho} (|g|^{1/4}\psi)$ in covariant form.

\section*{ACKNOWLEDGEMENTS}

I would like to thank I. Mikytiuk and E.~Tagirov for
useful discussions and M.~Gotay for pointing out the paper \cite{gotay}. 

\section*{APPENDIX}
\setcounter{equation}{0}
\def\theequation{A.\arabic{equation}}

We consider here the von Neumann rule (\ref{vN}). Relations of such type
appears in various quantization approaches (see e.g. \cite{GotGrTuy,vN}).
The simplest general form of von Neumann rule is
    \bqq{vN1}
{\cal Q} (\varphi \circ f) = \varphi ({\cal Q} f)
    \eeq
for a distinguished class of observables $C_D$, $f\in C_D$
and certain function $\varphi \in
C^\infty (\rr)$. It is easy to see that (\ref{vN1}) is not true for
all observables $C_D = C^\infty (N)$. However, it is obviously true
for restricted classes of functions on the phase space.

We shall prove here the relation
    \bqq{gvN}
{\cal Q} (\Phi (p_{\alpha_1},\ldots, p_{\alpha_k}) =
\Phi ({\cal Q} p_{\alpha_1}, \ldots, p_{\alpha_k})
    \eeq
where $\Phi$ is a symmetric $k$-linear real-valued mapping of the momentum
space.

Since $[\qq p_\alpha, \qq p_\beta]=0$ we have
    $$
[\qq (p_{\alpha_1} \ldots  p_{\alpha_k}), \qq p_\beta ]=
i\hbar\qq (\{ p_{\alpha_1} \ldots p_{\alpha_k}, \: \qq p_\beta \})=0
    $$
for any $\alpha_1 ,\ldots ,\alpha_k=1,\ldots, n$ because
$[\qq p_\alpha, \qq p_\beta]=0$. At the same time
    $$
[\qq (p_{\alpha_1} \ldots p_{\alpha_k}), \qq x^\beta ]=
i\hbar \qq (\{ p_{\alpha_1} \ldots p_{\alpha_k}, \qq x^\beta \})=
    $$
    $$
i\hbar\qq (\delta^\beta_{\alpha_1}\: p_{\alpha_2}\ldots p_{\alpha_k}+
\delta^\beta_{\alpha_k}\: p_{\alpha_1}\ldots p_{\alpha_{k-1}})=
[\qq (p_{\alpha_1}) \ldots \qq (p_{\alpha_k}), \qq x^\beta ].
    $$
From here it follows that $\qq (p_{\alpha_1} \ldots p_{\alpha_k})=
\qq (p_{\alpha_1}) \ldots \qq (p_{\alpha_k})+ T$ where $T$ is an operator
commuting with both momenta and coordinates.

Supposing by induction that (\ref{gvN}) holds for $k-1$and using the
formula \cite{gotay}
    $$
\qq (p_\alpha x^\beta ) = \frac{1}{2} (\qq (p_\alpha )\qq (x^\beta )
+ \qq ( x^\beta )\qq ( p_\alpha ))
    $$
we get
    $$
\qq (p_{\alpha_1} \ldots p_{\alpha_k}) =
\qq (\{ p_{\alpha_1} \ldots p_{\alpha_k}), p_\beta x^\beta\} )=
-\frac{i}{\hbar}[\qq (p_{\alpha_1} \ldots p_{\alpha_k}),
\qq (p_\beta x^\beta)]=
    $$
    $$
-\frac{i}{2\hbar} [\qq (p_{\alpha_1}) \ldots \qq (p_{\alpha_k}) + T,\;
\qq (p_\beta )\qq (x^\beta )
+ \qq ( x^\beta )\qq ( p_\beta )]=
\qq (p_{\alpha_1}) \ldots \qq (p_{\alpha_k}).
    $$
By linearity of the quantization mapping this proves the genralized
von Neumann rule (\ref{gvN}) and (\ref{vN}).

\def\theequation{\arabic{section}.\arabic{equation}}

\vskip1cm


\end{document}